\documentclass[trackchanges,twocolumn]{aastex701}

\usepackage{hyperref}
\usepackage{booktabs}   
\usepackage{siunitx}     
\usepackage{caption}     

\begin{document}

\title{Prevailing thermally-pulsing-asymptotic-giant-branch stars in the near-infrared rest-frame spectra of distant quiescent galaxies: towards robust galaxy ages and masses}

\author[orcid=0000-0001-5988-2202, sname='Shiying']{Shiying Lu}
\affil{School of Physics and Astronomy, Anqing Normal University, Anqing 246011, China} 
\affil{Institute of Astronomy and Astrophysics, Anqing Normal University, Anqing 246133, China}
\affil{Key Laboratory of Modern Astronomy and Astrophysics (Nanjing University), Ministry of Education, Nanjing 210093, China}
\email[show]{ShiyingLu@smail.nju.edu.cn} 
%\altaffiliation{Corresponding author: Shiying LU}

\author[orcid=0000-0002-3331-9590]{Emanuele Daddi} 
\affil{Universit{\'e} Paris-Saclay, Universit{\'e} Paris Cit{\'e}, CEA, CNRS, AIM, Paris} 
\email{emanuele.daddi@cea.fr}

\author[orcid=0000-0001-7711-3677]{Claudia Maraston} 
\affil{Institute of Cosmology and Gravitation, University of Portsmouth, Portsmouth, UK} 
\email{claudia.maraston@port.ac.uk}

\author[0000-0002-7093-7355]{Alvio Renzini} 
\affil{INAF-Osservatorio Astronomico di Padova, Padua, Italy} 
\email{alvio.renzini@oapd.inaf.it}

\author[0000-0002-6325-5671]{Daniel Thomas}
\affiliation{School of Physics and Astronomy, University of Leeds, Leeds, LS2 9JT, UK}
\email{D.G.Thomas@leeds.ac.uk}

\author[0000-0001-5414-5131]{Mark Dickinson}
\affil{NSF's National Optical-Infrared Astronomy Research Laboratory, 950 N. Cherry Ave. Tucson, AZ 85719, USA}
\email{mark.dickinson@noirlab.edu}

\author[0000-0002-7959-8783]{Pablo Arrabal Haro}
\affil{NSF's National Optical-Infrared Astronomy Research Laboratory, 950 N. Cherry Ave. Tucson, AZ 85719, USA}
\email{parrabalh@gmail.com}

\author[0000-0003-1586-2378]{Luis Gabriel Dahmer-Hahn}
\affil{Institute of Cosmology and Gravitation, University of Portsmouth, Portsmouth, UK}
\email{luisgdh@gmail.com}

\author[0000-0003-0121-6113]{Raphael Gobat}
\affil{Instituto de F{\'i}sica, Pontificia Universidad Cat{\'o}lica de Valpara{\'i}so, Casilla 4059, Valpara{\'i}so, Chile}
\email{raphael.gobat@pucv.cl}

\author[0000-0002-7831-8751]{Mauro Giavalisco}
\affil{University of Massachusetts Amherst, Amherst, MA, USA.}
\email{mauro@umass.edu}

\author[0000-0003-1581-7825]{Ray A. Lucas}
\affil{Space Telescope Science Institute, 3700 San Martin Drive, Baltimore, MD 21218, USA}
\email{lucas@stsci.edu}

\author[0000-0001-9879-7780]{Fabio Pacucci}
\affil{Center for Astrophysics $\vert$ Harvard \& Smithsonian, Cambridge, MA 02138, USA}
\affil{Black Hole Initiative, Harvard University, Cambridge, MA 02138, USA}
\email{fabio.pacucci@cfa.harvard.edu}

\author[0000-0003-3466-035X]{{L. Y. Aaron} {Yung}}
\affiliation{Space Telescope Science Institute, 3700 San Martin Drive, Baltimore, MD 21218, USA}
\email{yung@stsci.edu}

\author[0000-0002-3301-3321]{Michaela Hirschmann}
\affiliation{Institute of Physics, Lab for galaxy evolution and spectral modelling, EPFL, Observatory of Sauverny, Chemin Pegasi 51, 1290 Versoix, Switzerland}
\email{michaela.hirschmann@epfl.ch}

\author[0000-0002-4884-6756]{Benne Holwerda}
\affiliation{ Department of Physics and Astronomy, University of Louisville, Natural Science Building 102, Louisville, KY 40292, USA}
\email{bwholw01@louisville.edu}

\begin{abstract} 
We recently reported the discovery of prominent features from the thermally pulsing asymptotic-giant-branch (TP-AGB) phase in the rest-frame near-infrared of a massive quiescent galaxy (QG) at $z\sim 1$ observed with the JWST, which provides strong constraints on population synthesis { (SPS)} models. Here we { extend this analysis to 27} JWST/NIRSpec PRISM spectra of QGs at $z>1$ from GO-5019 and CEERS, with signal-to-noise ratios of $\sim$100 (15/27) and $\sim$50 (12/27), respectively. Each spectrum is modeled with three SPS models: the latest Maraston (M13) models with a sizable TP-AGB phase, and widely-used Bruzual \& Charlot 2003 (BC03) and Conroy \& Gunn 2009 (C09) models, both { with weaker} TP-AGB { contributions}. M13 generally provides the best { overall} fit { and the most consistency between the optical and the NIR. Only M13 yields consistent ages from separate fits to the optical, the NIR and the full wavelength ranges.} Compared to BC03 and C09, M13 yields systematically younger mass-weighted ages (by $<500$ Myr) hence lower stellar masses (by $<0.2$ dex). All models favor super-solar ($Z/Z_\odot >1.5$) metallicities. Signal-to-noise-weighted stacked spectra reveal that TP-AGB-related features are strongest in galaxies with mass-weighted ages of $t=0.4–1.8$ Gyr, consistent with the predicted peak TP-AGB contribution in M13. Further sample subdivisions show that these features are most pronounced in high-mass, dusty, and metal-rich systems. These results { support a significant contribution from} TP-AGB stars { to} the NIR spectra of high-redshift, intermediate-age galaxies and pave the way towards improved spectral population synthesis modeling and robust stellar ages and masses.
\end{abstract}

\section{Introduction}\label{intro}
\setcounter{footnote}{0}
Stellar population synthesis (SPS) models are essential for inferring the physical properties of galaxies from photometric and spectroscopic observations, including stellar age, metallicity, dust attenuation, the star formation history (SFH), and stellar masses \citep[e.g.][]{Bruzual+Charlot+03, Maraston+05, Conroy+09}. It is well known that capturing key physics in the synthetic models -- from stellar energetics and spectra, and in particular the treatment of the thermally pulsing asymptotic-giant-branch (TP-AGB) phase -- is key to avoid systematic uncertainties in the derived galaxy physical parameters.

\begin{figure*}[htpb]
    \centering
    \includegraphics[width=0.9\linewidth]{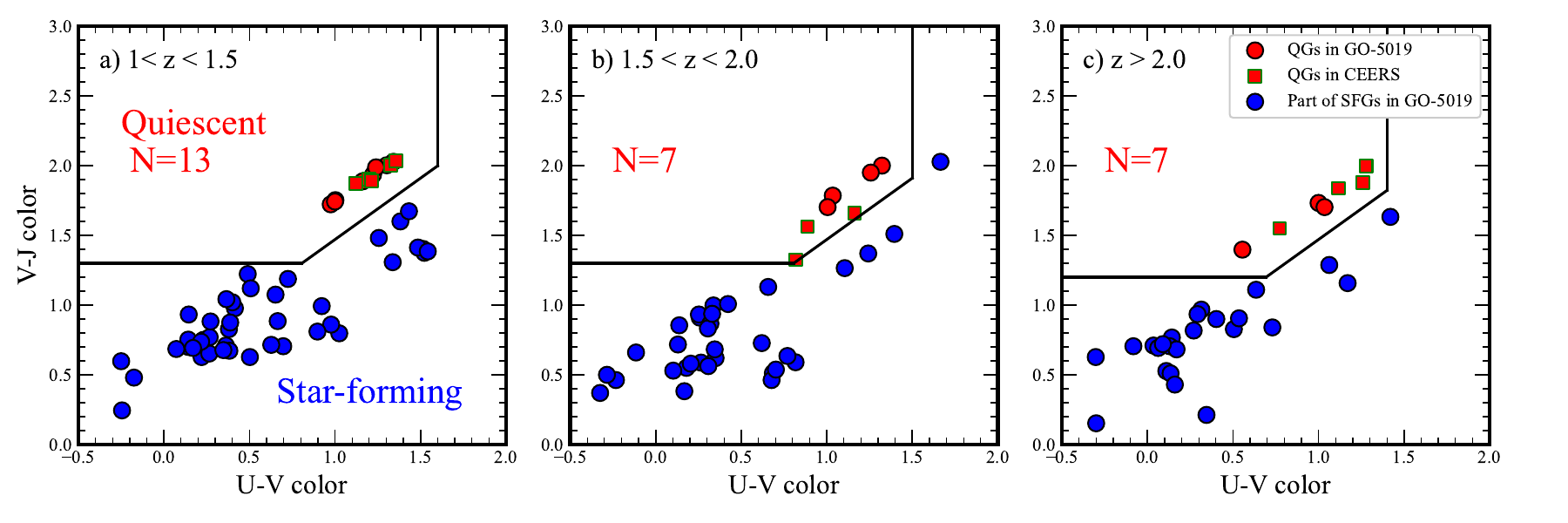}
    \caption{Rest-frame UVJ diagram for QGs at $z>$1. Black lines indicate the quiescent region from \citet{Whitaker+11}. QGs with spectra from GO-5019 (red circles) and CEERS (red squares) are labeled. Star-forming galaxies observed as fillers in GO-5019 are shown as blue circles in the same redshift bins.}
    \label{fig1_uvj}
\end{figure*}

TP-AGB stars dominate the { near-infrared} (NIR) light of intermediate-age stellar populations ($\sim$0.2-2 Gyr) in \cite{Maraston+1998, Maraston+05} evolutionary population synthesis models, producing numerous NIR features, including molecular absorption features (CN, TiO, C$_2$, CO) and the H-band bump at $\sim$1.6$\mu$m \citep[e.g.,][]{Verro+22, Liu+23, Lu+25}. Other population synthesis models (e.g., BC03, \citealt{Bruzual+Charlot+03}; C09, \citealt{Conroy+09}), include contributions from the TP-AGB phase of smaller magnitude than the former.  Ample literature over the past twenty years has demonstrated that the treatment of the TP-AGB stellar phase influences stellar ages and hence stellar masses that are derived when population synthesis models are used to interpret observed galaxy spectra and broad-band spectral energy distributions (SEDs, e.g., \citealt{van+der+Wel+05, Maraston+06, Capozzi+16}). In particular, models with a sizable TP-AGB phase such as M05 \citep{Maraston+05} and its latest revision with a lower energetic and a calibrated onset age (revised M05, hereafter M13; \citealt{Noel+13}) yield younger ages and lower stellar masses (by $\sim$0.1-0.3 dex) than light TP-AGB models such as BC03 \citep{Maraston+06, Capozzi+16}. This is because the NIR luminosity from TP-AGB stars allows the observed photometry to be reproduced with a younger age which then translates into a lower stellar mass \citep{Maraston+06}, over models where the same flux originates from older red-giant-branch (RGB) stars. 

Although these systematics have plagued the robust derivation of galaxy stellar ages and masses for two decades \citep{Maraston+05, Maraston+06, Conroy+09, Conroy+10, Kriek+10, Zibetti+13, Riffel+15, Capozzi+16}, a step forward towards a solution of this conundrum came from the detection of strong spectral features from TP-AGB stars in the rest-frame NIR { spectrum of the} massive quiescent galaxy { D36123 (signal-to-noise ratio, SNR$\sim$187)} at $z\sim1$ observed with JWST.
{ The comparison with} a wide range of SPS models { in \cite{Lu+25}} clearly { indicates} a sizable TP-AGB contribution in this galaxy, { with M13 models providing the best overall fits to the observed spectra.} The question then arises as to whether galaxy D36123 is an exceptional case, or whether this result is typical among the spectra of high-redshift { quiescent galaxies} (QGs)\footnote{QGs have the advantage that their spectra are dominated by evolved stellar phases rather than by star formation, but it should be pointed out that \cite{Riffel+07, Riffel+15} found evidence that including TP-AGB models provides better fits also in star-forming galaxies at low redshifts.}.

This Letter aims to address this question by analyzing JWST/NIRSpec PRISM observations of 27 QGs from program GO-5019 (PI: Shiying Lu) and the Cosmic Evolution Early Release Science (CEERS) survey (DD-ERS-1345; PI: S. L. Finkelstein, \citealt{Finkelstein+25}). We { therefore} adopt here three of the SPS models as in \cite{Lu+25}: M13, which we found to provide the best fit, { alongside BC03 and C09 that are routinely used in the community to derive galaxy properties, particularly at high redshift.} { We then fit the spectra, evaluate the goodness of fit, and compare the physical parameters derived for each model adopting the same procedures as \cite{Lu+25}, which allows us a straight comparison between the results.} 
Throughout this work, we adopt a flat $\Lambda$CDM cosmology with $H_0 = 70~{\rm km~s^{-1}~Mpc^{-1}}$, $\Omega_{\rm m} = 0.3$, and $\Omega_\Lambda = 0.7$. All magnitudes are reported in the AB system \citep{Oke+Gunn+1983}, and we assume the \cite{Chabrier+03} initial mass function.

\section{Sample and Data Reduction} \label{sec2}
\subsection{Quiescent Galaxy Sample from Program GO-5019 }
As part of JWST Cycle 3 observations, program GO-5019\footnote{ \url{https://mast.stsci.edu/portal/Mashup/Clients/Mast/Portal.html}} obtained NIRSpec multi-object spectroscopy using a single configuration of the Multi-Shutter Assembly (MSA) with the PRISM/CLEAR disperser/filter combination on 2025 April 19 (UTC), with a total exposure time of $\sim$2.55 hours, employing three-shutter slitlets per object and a three-point nodding pattern to facilitate background subtraction. The primary goal of this program is to substantially increase the sample of massive QGs at $1\lesssim z \lesssim 3$ with high-quality JWST/PRISM spectra in the Extended Groth Strip (EGS). The low redshift limit ensures that JWST/NIRSpec covers the age-sensitive spectral region around the rest-frame 4000\AA\ (i.e., D4000 break). The PRISM broad wavelength coverage (0.5-5.3 $\mu$m) encompasses numerous spectral features associated with AGB stars, which are critical for constraining the contribution of TP-AGB stars in SPS models. \cite{Lu+25} detected strong NIR spectral features from TP-AGB stars in the bright (Mag$_{\rm3.6\mu m}=$21.4) QG D36123. For this reason, for program GO-5019 we have selected QGs as bright as D36123 or brighter as primary targets with the highest priority, followed by fainter QGs with Mag$_{\rm3.6\mu m}<$23. A total of 18 QGs were selected using the UVJ color-color criteria (\citealt{Whitaker+11}, see Figure~\ref{fig1_uvj}), using colors obtained by \cite{Stefanon+17} from the  CANDELS photometric catalog \citep{Grogin+11, Koekemoer+11}.

A total of 16 out of the 18 primary QGs were successfully detected, with spectra covering the full wavelength range with average SNR$\sim$100. Spectra were also simultaneously secured for approximately 160 filler targets. These filler observations primarily target the nuclear regions of star-forming galaxies, which may host evolved bulges, offering complementary insights for future work. After visual inspection, we retained 15 primary QGs from GO-5019, excluding one target because of severe blend with a neighboring galaxy. The red circles in Figure~\ref{fig1_uvj} represent the 15 primary QGs from GO-5019. The blue circles represent fillers within our redshift intervals.

\subsection{Quiescent Sample from CEERS}
We also included QGs from the CEERS program, which released NIR spectroscopic observations for 1,325 galaxies in Data Release (DR) v0.7\footnote{\url{https://ceers.github.io/dr07.html}} and from the NIRSpec follow-up observations of program DD-2075 (PI: P. Arrabal Haro, \citealt{Arrabal_Haro+23_nat}). Although both programs are primarily designed to investigate the nature of (mostly) star-forming galaxies at very high redshift, 
the NIRSpec MSA slits serendipitously cover QGs as ``filler'' targets. 
To identify these QGs, we again applied the UVJ color-color selection criteria, by cross-matching the CANDEL EGS catalog \citep{Stefanon+17}. We selected a total of 17 QGs with mag$_{\rm 3.6\mu m}<$23 from the NIRSpec/PRISM CEERS observations. After visual inspection of individual spectra, we retained 12 QGs with an average SNR$\sim$50, which are shown as red squares in Figure~\ref{fig1_uvj}. The excluded galaxies include cases of poor spectral quality (1/17); the absence of a detectable D4000 break (2/17); or low SNR and featureless spectra that prevent secure redshift measurements (2/17). Together with the 15 QGs from GO-5019, our final combined sample includes a total of 27 QGs. 

\subsection{Reduction of the Spectral Data}
Data reduction for the NIRSpec spectra of the 27 QGs follows the same methodology adopted in previous CEERS NIRSpec studies \citep{Arrabal_Haro+23_apjl, Arrabal_Haro+23_nat, Lu+25}. The reduced CEERS spectra from DR v0.7 (Arrabal Haro et al., in prep.) have been processed with the JWST calibration pipeline \citep{Bushouse+22} version 1.8.5 provided by the Space Telescope Science Institute (STScI), using the Calibration Reference Data System (CRDS) context jwst\_1029.pmap. In contrast, the NIRSpec data from GO-5019 were reduced using a more recent version of the STScI calibration pipeline (v1.18.0) and CRDS context jwst\_1364.pmap. The complete reduction procedure is described in previous CEERS works \citep{Arrabal_Haro+23_apjl, Arrabal_Haro+23_nat, Lu+25}, for details see Appendix~\ref{AppendixA}. 

\begin{figure*}[htpb]
    \centering
    \includegraphics[width=1\linewidth]{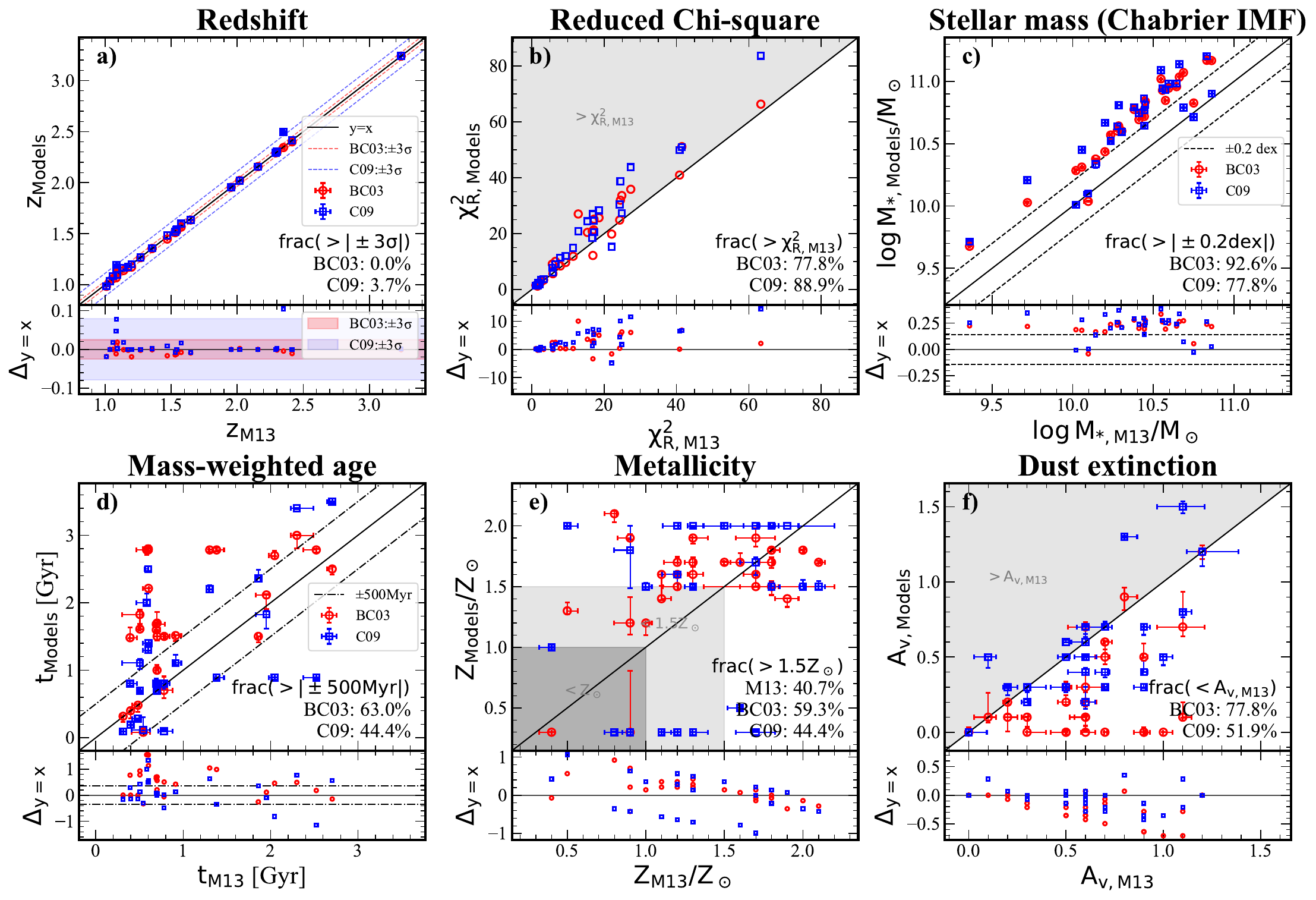}
    \caption{Comparison of galaxy parameters from M13 models (x-axis) and BC03 (red circles) or C09 (blue squares) models (y-axis), including the best-fit redshift ($z$, panel a), reduced chi-square ($\chi^2_{\mathrm{R}}$, panel b), stellar mass ($\log M_*/M_\odot$, panel c), mass-weighted age ($t$, panel d), metallicity ($Z/Z_\odot$, panel e), and dust extinction ($A_{\rm v}$, panel f). In each panel, the upper section shows the distributions of parameter values, while the lower section shows the distribution of the projected distances of individual data points to the one-to-one ($y=x$) relation. The fraction of sources satisfying the adopted selection criterion in models (relative to the physical limits shown in brackets) is shown in the lower-right corner of each panel. Shaded regions either mark where BC03/C09 yield higher $\chi^2_{\mathrm{R}}$ (panel b) and attenuation (panel f) than M13;  or metallicity $\le$1-1.5~$Z_{\odot}$ in panel {\it e}.}
    \label{fig2:pars_compare}
\end{figure*}

\section{Spectral Corrections and Fits} \label{sec3}
Multiwavelength photometric data in the CEERS field were released by \cite{Cox+25} as DR v1.0, including seven JWST/NIRCam filters spanning 1.15-4.44 $\mu$m and seven ancillary HST filters covering 0.435-1.6 $\mu$m. Since 17 galaxies, 63\% of the sample, lie within the CEERS/NIRCam footprint, we obtained photometry covering the range $\sim$0.6-4.44$\mu$m by cross-matching the CEERS DR v1.0 catalog. For galaxies outside the CEERS coverage, photometry was secured by cross-matching with the CANDELS EGS catalog \citep{Stefanon+17}, which includes data from HST, CFHT, and IRAC over the range $\sim$0.6-4.5$\mu$m. Following \cite{Lu+25}, we applied an aperture correction { to each individual spectrum (see Appendix~\ref{AppendixB}),} based on the available multi-band photometry to account for the actual shape of each target. This assumes that no strong color gradients are present between the position of the slit and the rest of the photometric aperture.  
After this initial correction, the PRISM spectrum is further corrected for the total galaxy-integrated SED (see Figure~\ref{ext_figB:apr_corr}).

Before performing the spectral fitting, we accounted for the relevant sources of spectral broadening, including the JWST/PRISM instrumental resolution, the intrinsic velocity dispersion of early-type galaxies, and differences in template resolution arising from spectral sampling. The effective spectral resolution may also depend on the source size, but the predominantly compact light profiles of our QGs minimize this effect. For the JWST/PRISM resolution, \cite{de_Graaff+24} reported that the in-flight spectral resolution is higher than the pre-launch estimate by a factor of $\sim$1.5-2.0. This is consistent with \cite{Lu+25}, who measured an increase of $\sim$1.4 for two QGs (MPT-IDs=D36123 and 8595) observed with PRISM. We therefore adopt a conservative multiplicative factor of $\times$1.4 to improve the PRISM resolution { (see Figure~\ref{ext_figC:res_ex})}. We account for the combined effect of intrinsic galaxy velocity dispersion and template spectral resolution, as in \cite{Lu+25} (see Appendix~\ref{AppendixC}).

For each QG spectrum, we then perform spectral fitting using M13, BC03, and C09 models. These models differ primarily in their treatment of TP-AGB stars, leading to systematic variations in the predicted NIR continuum shape and absorption features. Fits are performed with the same custom IDL code as in \cite{Lu+25}. This code assumes delayed-$\tau$ SFHs and corrects for dust reddening to find model combinations minimizing $\chi^2_{\rm R}$ (see Appendix~\ref{AppendixD} { and Table~\ref{ex_tabG:SSP_grid}}). { Several robustness tests, including alternative SPS models and dust attenuation prescription, and an independent spectral fitting code, are presented in Appendix~\ref{AppendixE}.}

\begin{figure*}[htpb]
    \centering
    \includegraphics[width=1\linewidth]{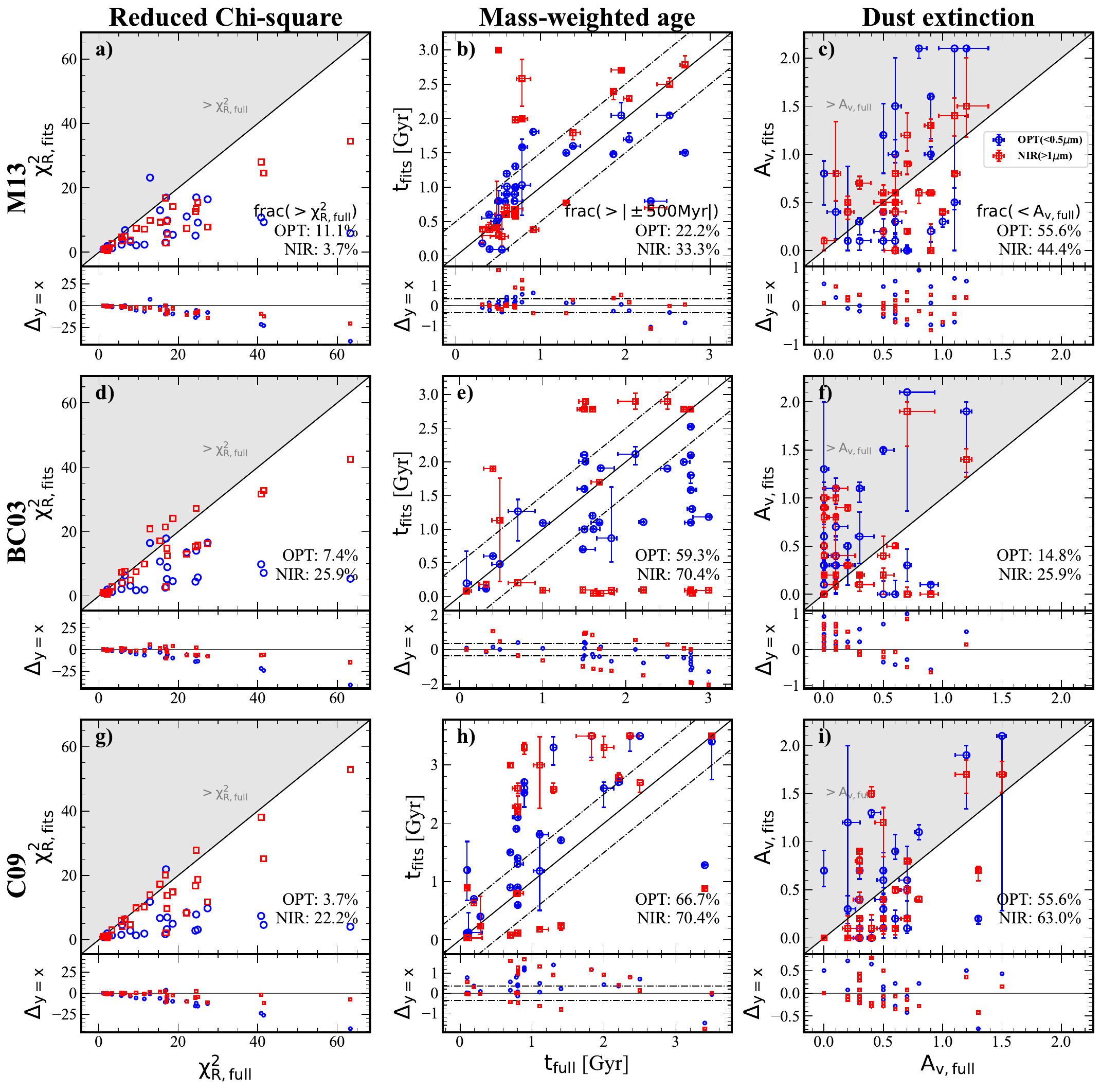}
    \caption{Comparison of results from full-spectrum fitting and from optical-spectrum ($\lambda_{\rm rest}$$<$0.5$\mu$m; blue circles) or NIR-spectrum ($\lambda_{\rm rest}$$>$1$\mu$m; red squares) only, for different SPS models, namely: M13 (top row); BC03 (middle row); C09 (bottom row). Left, middle, and right panels show $\chi^2_{\rm R}$, mass-weighted age, and dust extinction, respectively. In the optical-only and NIR-only fits, metallicity was fixed to the best-fitting value derived from the full-spectrum fit for each model.  Other reference lines, annotations, and plotting conventions are the same as in Figure~\ref{fig2:pars_compare}.}
    \label{fig3:pars_compare_fixmet}
\end{figure*}

\begin{figure*}[htpb]
    \centering
    \includegraphics[width=0.9\linewidth]{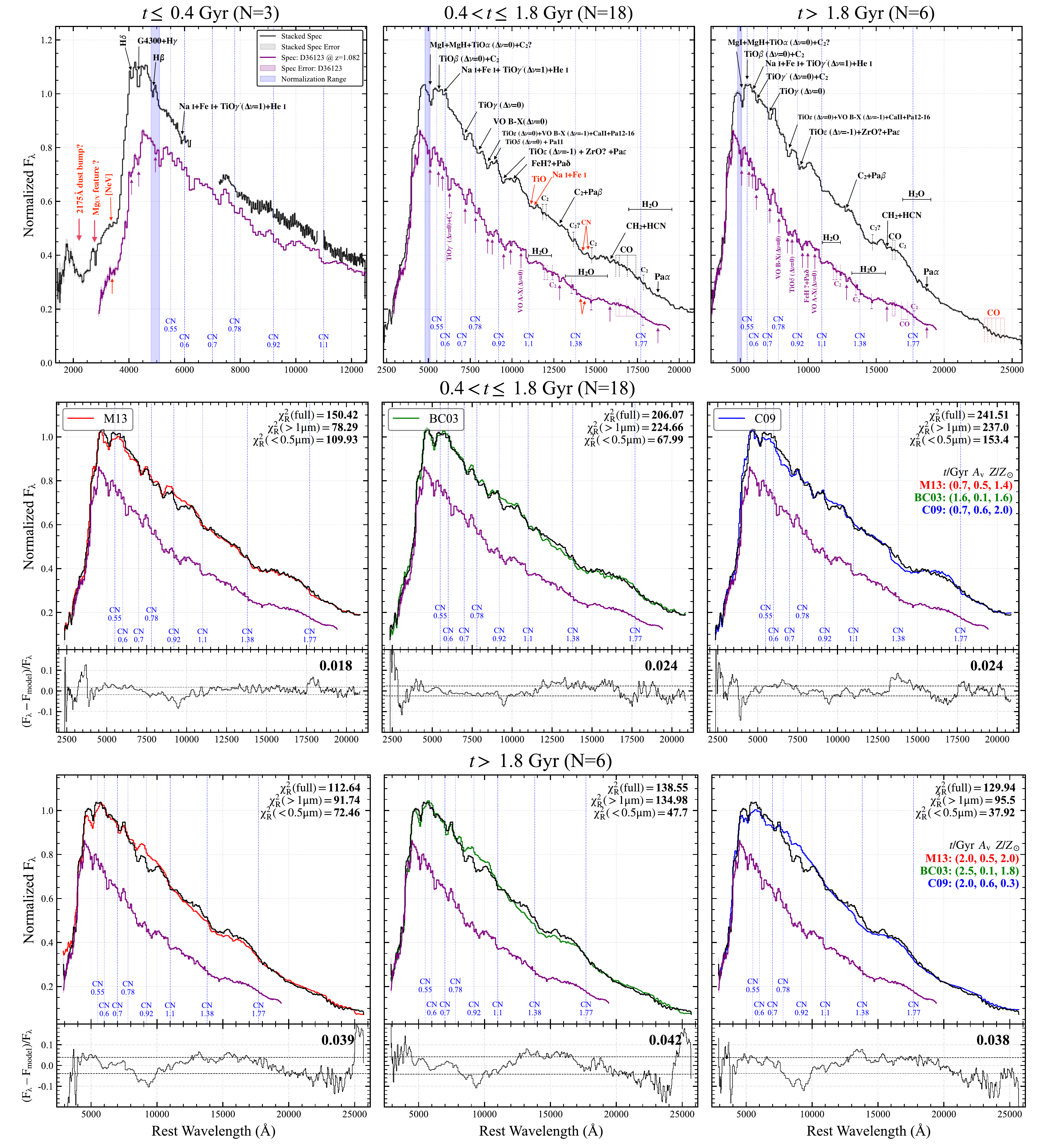}
    \caption{Stacked spectra (top row) and model fits (middle and bottom rows) of quiescent galaxy (QG) subsamples binned by mass-weighted age ($t$). The top row (left to right) shows stacks for $t \le 0.4$ Gyr, $0.4 < t \le 1.8$ Gyr, and $t > 1.8$ Gyr, with sample sizes labelled at the top of each panel. Black histograms show the stacked spectra with uncertainties (gray shading). In the younger bin ($t \le 0.4$ Gyr), masked regions in the stacked spectrum appear blank. The spectrum of D36123 \citep{Lu+25} is shown in purple for comparison. All spectra are normalized to the median flux in the rest-frame 4800-5100 \AA\ window (blue shaded region). NIR absorption features from \cite{Lu+25} are labeled (black if matched; purple if unmatched), with CN edges marked by blue dashed lines. Additional new identifications are marked in red. The top sub-panels in the middle and bottom rows show the best-fit model for M13 (red), BC03 (green), and C09 (blue), respectively, in the intermediate ($0.4 < t \le 1.8$ Gyr) and older ($t > 1.8$ Gyr) age stacks, as labeled. Reduced $\chi^2$ values (full, NIR with $\lambda_{\rm rest} > 1\mu$m, { and optical with $\lambda_{\rm rest} < 0.5\mu$m}) and best-fit parameters are also shown. The bottom sub-panels of the middle and lower rows show relative flux deviations, $D_{\rm REL} = (F_\lambda - F_{\rm model})/F_\lambda$, with $\sigma_{D_{\rm REL}}$ indicated by dashed lines and SNR-weighted averages labeled at the top right.}
    \label{fig4:stack_fit}
\end{figure*}

\section{Results} \label{sec4}
\subsection{Physical Parameters from Individual Spectra} \label{sec4.1}
\subsubsection{Comparison of full-spectrum fits} \label{sec4.1.1}
In Figure~\ref{fig2:pars_compare}, we compare galaxy properties derived from BC03 (C09) models with those obtained from M13 models { (listed in Table~\ref{ext_tabD:list})}. All three SPS models produce consistent spectroscopic redshifts, as nearly all measurements fall well within the 3$\sigma$ deviation from the one-to-one relation (panel a), demonstrating that for our sample, redshift estimates are robust against the choice of SPS models.

Assessing the overall fitting performance, we find that in 78\% (89\%) of the galaxies, M13 models yield best-fit solutions with lower reduced Chi-square values, $\chi^2_{\rm R}$, than BC03 (C09) models (panel b), consistent with previous results (e.g., \citealt{Capozzi+16, Lu+25}). Systematic differences are evident in the stellar masses derived with TP-AGB-light models (BC03 and C09, in Figure~\ref{fig2:pars_compare}c), which are systematically higher by $\sim$0.2 dex for 93\% (BC03) and 78\% (C09) of their samples, respectively. This offset stems from the older galaxy ages obtained with TP-AGB-light models (see Figure~\ref{fig2:pars_compare}d), where the NIR flux required to reproduce the data is provided by RGB stars whose contribution becomes dominant at an age of $\sim$1 Gyr \citep{Greggio+11}. In turn, older ages imply higher inferred stellar masses, as well known. 

Age (mass-weighted, $t$, panel d), metallicity (panel e), and dust attenuation ($\rm A_{v}$, panel f) are degenerate parameters in spectral fitting. Figure~\ref{fig2:pars_compare} reveals systematic differences between the various SPS models. M13 yields ages around $\sim$0.6 Gyr for about half of the galaxies, whereas BC03 and C09 models spread their ages up to $\sim$2.5 Gyr, well into the RGB-dominated epoch of stellar populations. { M13 generally yields higher best-fitting A$_{\rm v}$ values than BC03 and C09 (Figure~\ref{fig2:pars_compare}f), consistent with the younger ages inferred by M13. In contrast, TP-AGB-light models preferentially converge} to older, RGB-dominated { stellar populations with lower A$_{\rm v}$}.  { Fixing A$_{\rm v}=$0 for all galaxies preserves the same relative trends in ages and stellar masses, demonstrating that these model-dependent differences are robust against the adopted dust attenuation (see Appendix~\ref{AppendixE.2}).} Notably, for over half of the sample, all SPS models yield super-solar metallicities, in many cases with $Z/Z_\odot>$1.5. This compares to $Z/Z_\odot \sim$1.2 for the stacked spectrum of a sample of QGs at $z\sim$1.6 \citep{Onodera+15}.

\subsubsection{Full-spectrum versus optical-only and NIR-only fits} \label{sec4.1.2}
{ To assess the wavelength dependence of the fits, we repeated the spectral fitting using only the rest-frame optical\footnote{Our adopted optical range is shorter than the conventional $\sim$0.7$\mu$m, chosen to minimize TP-AGB contributions, which are negligible below 0.5$\mu$m.} ($\lambda_{\rm rest}$$<$0.5$\mu$m) or rest-frame NIR ($\lambda_{\rm rest}$$>$1$\mu$m), while fixing the metallicity to the value from the full-spectrum best-fit (see Figure~\ref{fig3:pars_compare_fixmet}). Optical-only fits yield lower $\chi^2_{\rm R}$ values for all SPS models, indicating that optical spectra are similarly well reproduced, while the main model differences arise in the TP-AGB-sensitive NIR region. Notably, only M13 provides self-consistent mass-weighted ages in optical, NIR, and full-spectrum fits, whereas BC03 and C09 show systematic offsets toward younger and older ages, respectively, for the optical-only fits, and large noise for the NIR-only ages.}

\subsection{Analysis of Stacked Spectra} \label{sec4.2}
\subsubsection{The stacking procedure} \label{sec4.2.1}
The high quality of the JWST/PRISM spectra enables us to probe relatively weak TP-AGB features in the NIR via stacking. Before stacking, each spectrum was visually inspected to mask residual artifacts (e.g., skyline contamination and bad pixels). To avoid blurring NIR features in the stacks, individual spectra are shifted to the rest frame adopting redshifts ($z_{\rm spec}^{\rm adopted}$ in Table~\ref{ext_tabD:list}) that are robust, which are defined by requiring the normalized redshift difference between models to be $|z_{\rm M13} - z_{\rm model}|/(1 + z_{\rm M13}) < 0.002$ for both BC03 and C09, or that are well matched to unambiguous NIR features (e.g., CO through visual inspection). We apply $\lambda_{\rm rest} = \lambda_{\rm obs}/(1+z)$ and scale the $F_\lambda$ flux density by $(1+z)$ to conserve the integrated flux. The associated uncertainties were transformed accordingly. Each spectrum was resampled onto a common wavelength grid defined by the highest SNR spectrum in the selected sample, preserving its native sampling. A flux-conserving rebinning algorithm (``drizzing'') was implemented by computing fractional overlaps between input and output wavelength bins. Fluxes were redistributed using the overlap-weighted means, and the uncertainties were propagated in quadrature with the same weights. Wavelength regions not covered by a spectrum were assigned zero flux and zero weight so that they do not affect the stack. Each spectrum was normalized by the median continuum flux within the rest-frame 4800-5100\AA\ window, chosen for its broad coverage and minimal contamination by strong features in most sources. When necessary, the continuum level was estimated via linear interpolation across nearby uncontaminated regions. The final stacked spectrum was constructed using inverse-variance weighting ($w_i=1/\sigma_i^2$), with the corresponding spectral resolution treated using the same method { over 2-pixel resolution elements}.

\begin{deluxetable*}{c|c|c|c|c|c}[ht]
\digitalasset
\tablewidth{0pt}
\tablecaption{Specific features measurements for stacked spectra ($t>0.4$Gyr) in different bins \label{tab1:Features}}
\tablehead{
\colhead{Subsample} & \colhead{Stacked bin}  & \colhead{I(H-bump)} & \colhead{I(CN1.1)}& \colhead{I(CN0.92)}& \colhead{I(CN0.78)} \\
\colhead{(1)} & \colhead{(2)} & \colhead{(3)} & \colhead{(4)} & \colhead{(5)} & \colhead{(6)}}
\startdata
&  total    & -0.079$\pm$0.003 & -0.052$\pm$0.001 & -0.109$\pm$0.001 & -0.096$\pm$0.001 \\
\cline{2-6}
&  M13$^a$      & -0.080(0.016) & -0.057 & -0.060 & -0.091 \\
&  BC03$^a$     & -0.088(0.029)  & -0.035 & -0.043 & -0.091\\
 &  C09$^a$      & -0.120(0.030) & -0.014& -0.038 & -0.061 \\
\cline{2-6}
&  $\log M_*/M_\odot\ge$10.445  & -0.087$\pm$0.003 & -0.097$\pm$0.001 & -0.109$\pm$0.001 & -0.087$\pm$0.001 \\
TP-AGB dominated & $\log M_*/M_\odot<$10.445   & -0.071$\pm$0.003 & -0.084$\pm$0.002 & -0.104$\pm$0.002 & -0.065$\pm$0.002 \\
\cline{2-6}
($t=$0.4-1.8 Gyr)& A$_{\rm v}\ge$0.6 &-0.085$\pm$0.002 & -0.081$\pm$0.002 & -0.113$\pm$0.001 & -0.084$\pm$0.001 \\
& A$_{\rm v}<$0.6 &-0.072$\pm$0.003 & -0.048$\pm$0.001 & -0.111$\pm$0.001 & -0.083$\pm$0.001 \\  
\cline{2-6}
& $Z/Z_\odot\ge$1.35 & -0.070$\pm$0.002 & -0.084$\pm$0.001 & -0.115$\pm$0.001  & -0.108$\pm$0.001\\
& $Z/Z_\odot<$1.35& -0.094$\pm$0.003 & -0.077$\pm$0.001 & -0.107$\pm$0.001  & -0.079$\pm$0.001 \\
\hline
older  ($t>$1.8Gyr) & total & -0.087$\pm$0.003 & -0.043$\pm$0.002 & -0.072$\pm$0.002 & -0.090$\pm$0.002 \\
\cline{2-6}
&  M13$^a$      & -0.073(0.036) & -0.048 & -0.062 & -0.068 \\
&  BC03$^a$     & -0.081(0.037)  & -0.030 & -0.034 & -0.062\\
&  C09$^a$      & -0.078(0.031) & -0.020& -0.040 & -0.030\\
\enddata
\tablecomments{Column 3 gives the strength of the H-band bump at $\sim$1.6 $\mu$m, and Columns 4-6 report the strength of the CN absorption edge at 1.1, 0.92, and 0.78 $\mu$m (see Appendix~\ref{AppendixF}). Spectral fits are performed on stacked spectra of the TP-AGB–dominated and older subsamples for M13/BC03/C09 models, and the corresponding spectral features are also measured. Measurements are performed following similar definitions and methodology described in \cite{Lu+25}. Superscript $^a$ denotes SNR-weighted averages of the relative flux deviation, $D_{\rm REL} = (F_\lambda - F_{\rm model})/F_\lambda$, given in parentheses.}
\end{deluxetable*}

\subsubsection{Model fits and spectral features in stacks}\label{sec4.2.2}
Following the \cite{Noel+13} calibration,  we divide our QG sample into three bins of mass-weighted age from M13 best fits: $t\le 0.4$ Gyr ($N=$3), the $0.4 < t \le 1.8$ Gyr ($N=$18, { near-peak TP-AGB contribution}), and $t>1.8$ Gyr ($N=$6). Using the stacking procedure described in Section~\ref{sec4.2.1}, we construct stacked spectra for each age bin. Note that all three M13-age bins are likely to include some TP-AGB contribution, given the relatively broad distribution of stellar ages implied by the adopted, delayed-$\tau$ SFH. The stacked spectra for each age bin are shown as black histograms in Figure~\ref{fig4:stack_fit}. Table~\ref{tab1:Features} reports the measurements of selected absorption features on the stacks, whose strength is sensitive to TP-AGB stars, namely the H-bump and three CN band-heads at various wavelengths (1.1, 0.92, and 0.78$\mu$m, { see Appendix~\ref{AppendixF} for measurements}). The H-band bump at $\sim$1.6$\mu$m is sensitive to cool evolved stellar populations. We measure it following \cite{Lu+25}, but with improved continuum fitting and integration over 1.5-1.8 $\mu$m. 
 
Compared to the TP-AGB-dominated bin { ($0.4<t \le 1.8$ Gyr)}, the stacked spectrum of the older subsample ($t>1.8$ Gyr) exhibits a relatively featureless NIR continuum, with fewer TiO absorptions (black labels in the top panels of Figure~\ref{fig4:stack_fit}) and weaker CN edges (also see Table~\ref{tab1:Features}), consistent with negligible contributions from intermediate-age TP-AGB stars. The youngest bin ($t\le0.4$ Gyr) shows a steeper continuum with numerous weak absorption features, likely originating from young to intermediate-age stars, including red supergiants, but still lacks the strong CN features seen in the TP-AGB-dominated bin and in D36123 \citep{Lu+25}. In contrast, the spectrum of the TP-AGB-dominated bin exhibits pronounced CN absorption features (see Table~\ref{tab1:Features}), confirming that TP-AGB stars contribute most strongly at intermediate ages.

Best-fit models for the stacked spectra are shown in the top sub-panels of the second and third rows in Figure~\ref{fig4:stack_fit} (see Figure~\ref{ext_figG1:stack_fit_lt0p4} for $t \le 0.4$ Gyr). For the TP-AGB-dominated bin, BC03 gives older ages ($t\sim$1.6Gyr) than M13/C09 ($t\sim$0.7Gyr), while all models return super-solar metallicities ($Z/Z_\odot > 1.4$), consistent with the results in Section~\ref{sec4.1.1}.  M13 yields the lowest $\chi^2_{\rm R}$ in the full range, providing the best overall fits. { This advantage is most pronounced at $\lambda_{\rm rest}>1\mu$m, where the $\chi^2_{\rm R}$ is} 78 compared to 225 for BC03 and 237 for C09. { In the short $\lambda_{\rm rest} < 0.5\mu$m, where TP-AGB-sensitive features are substantially weaker, BC03 has lower $\chi^2_{\rm R}$ values than M13 in both bins, while C09 in the older bin only.} 

The H-band bump at $\sim$1.6$\mu$m and the CN features at 1.1 and 0.78 $\mu$m are better reproduced by M13 (see Table~\ref{tab1:Features}). None of the models can fit the absorption feature coincident with the $\lambda_{\rm rest}=$0.92$\mu$m CN bandhead, and the failure is even more pronounced in the older age bin for BC03 and C09. { We tested that an enhanced Carbon abundance in population models does increase CN band-heads equivalent widths, both using ALF ([C/Fe]=0.15, \citealt{Conroy+18}) and our new models with [C/Fe] up to 0.5 (Maraston et al., {\it in prep.}). However, we also find that an enhanced [C/Fe] alone is not sufficient and that a TP-AGB contribution as in M13 needs to be included alongside. Detailed spectral modelling including abundance effects and spectral type mixing is the subject of a forthcoming paper.} 
All results are insensitive to the adoption of different SFHs, attenuation laws, { and alternative age-bin definitions for the stacked spectra based on BC03- or C09-derived ages} (see Appendix~\ref{AppendixG} for details).

\subsubsection{Inferred properties from the TP-AGB-dominated stacks} \label{sec4.2.3}
Compared to the young ($t\le$0.4 Gyr) and old ($t>$1.8 Gyr) stacks, the TP-AGB-dominated stack ($0.4 < t \le 1.8$ Gyr) exhibits significantly stronger TP-AGB-related spectral features. To investigate whether the strength of these features correlates with galaxy physical properties -- namely stellar mass ($\log M_*/M_\odot$), dust attenuation (A$_{\rm v}$), and metallicity ($Z/Z_\odot$) -- we further subdivide the TP-AGB-dominated subsample into two bins of comparable size based on each parameter { (see Figure~\ref{ext_figG2:stack_2bins})}.

By comparing three strong CN absorption troughs (quantified in Table~\ref{tab1:Features}), we find that { most} TP-AGB features are more prominent in galaxies with higher stellar masses ($\log M_*/M_\odot \ge$10.445), stronger dust attenuation ($\rm A_{v} \ge 0.6$), and higher metallicities ($Z/Z_\odot \ge 1.35$). Measurements in Table~\ref{tab1:Features} show that more massive galaxies with higher dust attenuation { probably} exhibit stronger H-band bumps, which M13 reproduces better than the other models (see also Figure~\ref{fig4:stack_fit}). In the intermediate-age regime ($t$=0.4-1.8 Gyr), TP-AGB stars reach cool effective temperatures and develop strong molecular absorption bands (e.g., CN and TiO), thereby showing enhanced NIR spectral features relative to both younger and older populations. The strength of these molecular features is also expected to depend on metallicity and specific element abundances, with more metal-rich stellar populations having stronger oxygen-rich molecular absorption in integrated light \citep{Maraston+05}, leading to the identification of more TiO features (see Figure~\ref{fig4:stack_fit}). Therefore, the observed trends likely reflect covariant effects (including the age-$Z$-A$_{\rm v}$ degeneracy) rather than a single dominant physical driver.

\section{Summary}\label{sec5}
We present the analysis of JWST/NIRSpec PRISM spectra of 27 high-redshift ($1<z\lesssim3$) QGs to probe the spectral signatures of the TP-AGB stellar phase in the NIR region. This is a follow-up study of our detection of strong AGB spectral features in the JWST spectra of a massive quiescent galaxy at $z\sim 1$ \citep{Lu+25}. { By comparing these new spectra with those from our previous work \citep{Lu+25}, we find that TP-AGB stars are ubiquitous in the spectra of QG at Cosmic Noon, confirming our previous results based on one single galaxy.} 

By fitting individual as well as stacked spectra with three { SPS} models (M13, BC03, and C09), We find that the M13 models with a sizable TP-AGB contribution { generally} provide better { overall} fits (lower $\chi^2_{\rm R}$ values), especially in the NIR, { and are the only models yielding self-consistent ages from the optical, NIR, and full-spectrum fits. Compared to BC03 and C09, M13 also yields systematically} younger mass-weighted ages (by 500 Myr) and lower stellar masses (by $\sim$0.2 dex), { while all} models consistently point to super-solar metallicities for about half of the sample.

The SNR-weighted stacked spectra reveal enhanced CN and H-bump features alongside many weaker TiO absorptions in galaxies with mass-weighted ages of 0.4-1.8 Gyr, consistent with the expected peak contribution of TP-AGB stars from M13 in this age range. The strength of these features { likely} increases toward higher stellar mass ($\log M_*/M_\odot \ge 10.445$), stronger dust attenuation ($\rm A_v \ge$0.6), and higher metallicity ($Z/Z_\odot \ge 1.35$). These results provide further direct empirical constraints to TP-AGB prescriptions in SPS models and underscore the importance of modelling evolved stellar phases to interpret the NIR spectra of distant QGs as available in the JWST era.

{ Our results suggest that the treatment of the TP-AGB phase remains an important factor of systematic uncertainty for deriving the physical properties of high-redshift QGs. Compared to M13, TP-AGB-light models tend to infer older ages hence higher stellar masses, with potential implications for galaxy formation studies. At the same time, no current SPS model reproduces all NIR spectral features fully, motivating further improvements, including elemental abundance variations (e.g., carbon and $\alpha$ enhancement) and spectral type variations. We shall pursue these avenues in a forthcoming paper.}

%% Please use the acknowledgment and contribution environments. This will 
%% be anonomyized when the "anonymous" style option is used. 
\begin{acknowledgments}
{ We thank the anonymous referee for the constructive comments that improved the paper.}
This work is supported by the National Natural Science Foundation of China (No. 12503011) and the UK Research and Innovation Science and Technology Grant UKRI1197 (PI: Claudia Maraston). S.L. acknowledges the support from the Key Laboratory of Modern Astronomy and Astrophysics (Nanjing University) by the Ministry of Education and { the support from the Program for Innovative Research Team in Anqing Normal University}. M.D. and P.A.H. acknowledge support from NASA through the Early Release Science Program of the Space Telescope Science Institute (Award JWST-ERS-1345) and the JWST-GO-2750 award. R.G. acknowledges funding from program ANID Fondecyt 1231661. This work is based on observations with the NASA/ESA/CSA JWST obtained from the Mikulski Archive for Space Telescopes at the Space Telescope Science Institute, which is operated by the Association of Universities for Research in Astronomy, Incorporated, under NASA contract NAS5-03127.
\end{acknowledgments}

\begin{contribution}
S.L. conceived the project, selected and identified the galaxies, and performed the raw data reduction of the GO-5019 observations. S.L., E.D., and C.M. led the analysis and wrote the manuscript. A.R., D.T., L.D.H., and M.G. assisted in the analysis and interpretation. P.A.H. and M.D. led the original observations and reduced the NIRSpec spectra from CEERS. R.G. wrote the fitting code. All authors aided in the analysis and interpretation and contributed to the final manuscript.
\end{contribution}

\facilities{JWST, HST, CFHT, IRAC}

\appendix
\section{Data reduction and example spectra }\label{AppendixA}
The primary QG sample from program GO-5019 was reduced following the same methodology of \cite{Arrabal_Haro+23_apjl, Arrabal_Haro+23_nat}, with a detailed description in preparation (Arrabal Haro et al.).  It includes three primary pipeline stages. In brief, the \textit{calwebb\_detector1} module converts uncalibrated detector images into count-rate maps, which serve as input for subsequent processing. The \textit{calwebb\_spec2} stage applies background subtraction, flat-field correction, wavelength and flux calibration, and two-dimensional (2D) distortion correction. Finally, \textit{calwebb\_spec3} produces the combined 2D and extracted one-dimensional (1D) spectra for sources MSA observed. In this work, the final 1D spectra were extracted using optimized apertures, with uncertainties computed by the JWST pipeline based on instrumental noise. We rescale these uncertainties by a factor of $\sim$1.4 to account for correlations introduced by the pipeline \citep{Arrabal_Haro+23_apjl, Lu+25}.

Example 2D (top) and 1D (bottom) spectra of QGs in program GO-5019 are shown in the right panel of Figure~\ref{ext_figA:ex_2D1D}, where the collapsed spatial light profile (top-right) illustrates the derivation of the optimal 1D spectrum. The top 2D spectra show the spatial region used for extraction (solid red line) and the 2$\sigma$ limits of the Gaussian kernel (dashed red line). Briefly, following the prescription of \cite{Horne+1986}, the optimal extraction weights the flux by the modeled spatial profile. To avoid contamination from multiple peaks (e.g., due to more than one source within a slitlet) in the 2D spectra, the extraction is restricted to a spatial window (blue line) centered on the expected target position, with median filtering along the spectral direction (purple line) applied to suppress image defects.

\setcounter{figure}{0} 
\renewcommand{\thefigure}{A\arabic{figure}}
\begin{figure*}[htpb]
    \centering
    \includegraphics[width=1\linewidth]{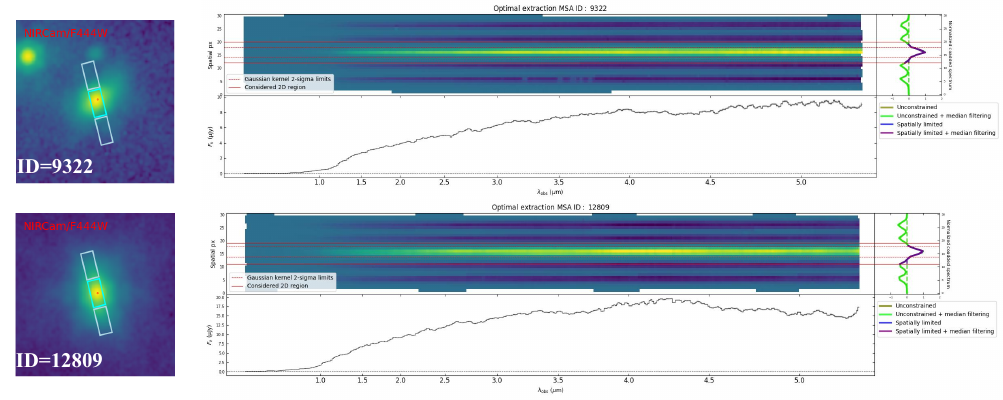}
    \caption{Sample NIR images (left) and 2D/1D spectra (right) for targets in the primary QG sample from GO-5019. The NIRSpec shutters used for detection are overlaid on the NIRCam/F444W images. In the right panels, the top shows the 2D spectrum with the optimal extraction aperture (red dashed line) derived from the best fit on the normalized coadded spectrum, while the bottom shows the corresponding observed 1D spectrum.}
     \label{ext_figA:ex_2D1D}
\end{figure*}

\section{Aperture correction} \label{AppendixB}
To obtain the galaxy-integrated spectrum, we perform an aperture correction for each target using all available HST and JWST \citep{Cox+25} photometry. Otherwise, available photometry \citep{Stefanon+17} from HST, CFHT, and IRAC is adopted. Briefly, for galaxy ID=9322, we first compute the synthetic photometry ($F_{\rm syn}$) by convolving the observed spectrum with the filter bandpasses. The aperture correction factor, $R_{\rm corr}$, is then derived as the ratio of the imaging photometry ($F_{\rm phot}$) to the synthetic photometry ($F_{\rm syn}$). Based on the best linear fit of $R_{\rm corr}$ as a function of wavelength (red line in the middle panel of Figure~\ref{ext_figB:apr_corr}), the observed spectrum of galaxy ID=9322 is rescaled to match the multi-band photometry (blue squares; left and right panels).
{ \cite{Lu+25} also derived the aperture correction using the JWST/NIRSpec Multi-Shutter Array forward-modeling tool (MSAFIT; \citealt{de_Graaff+24}), which accounts for the galaxy morphology, the NIRSpec point spread function, and detector pixelation. They showed that both alternative aperture-correction prescriptions and the $\pm1\sigma$ uncertainty have a negligible impact on the spectral fits and the derived stellar population parameters. As the treatment of aperture-correction uncertainties and their correlations is not yet well established, we do not add them in quadrature. In any case, this would only change the absolute $\chi^2$ values, not the relative ranking of the models, and therefore would not affect our conclusions.}

\setcounter{figure}{0} 
\renewcommand{\thefigure}{B\arabic{figure}}
\begin{figure*}[htpb]
    \centering
    \includegraphics[width=0.9\linewidth]{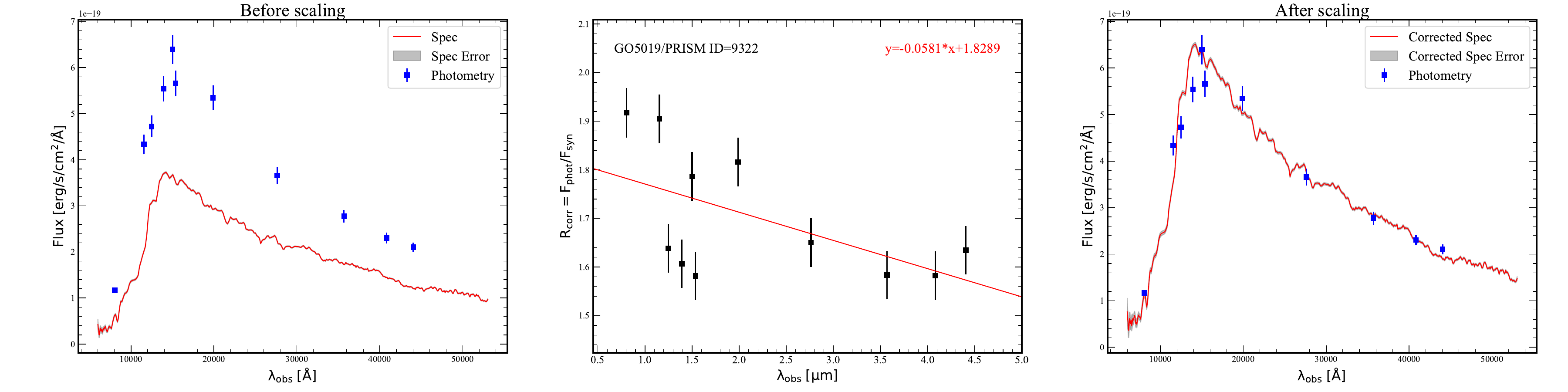}
    \caption{Example of the aperture correction for galaxy ID = 9322. The left and right panels show the spectrum (red) before and after aperture correction, compared with the observed multi-band photometry (blue squares). The middle panel shows the aperture correction factor (R$_{\rm corr}$) as a function of different observed bands, derived from the ratio of the measured imaging photometry (F$_{\rm phot}$) to the synthetic photometry (F$_{\rm syn}$). The best-fit relation and corresponding function are shown in red.}
    \label{ext_figB:apr_corr}
\end{figure*}

\setcounter{figure}{0} 
\renewcommand{\thefigure}{C\arabic{figure}}
\begin{figure*}[htpb]
    \centering
    \includegraphics[width=0.9\linewidth]{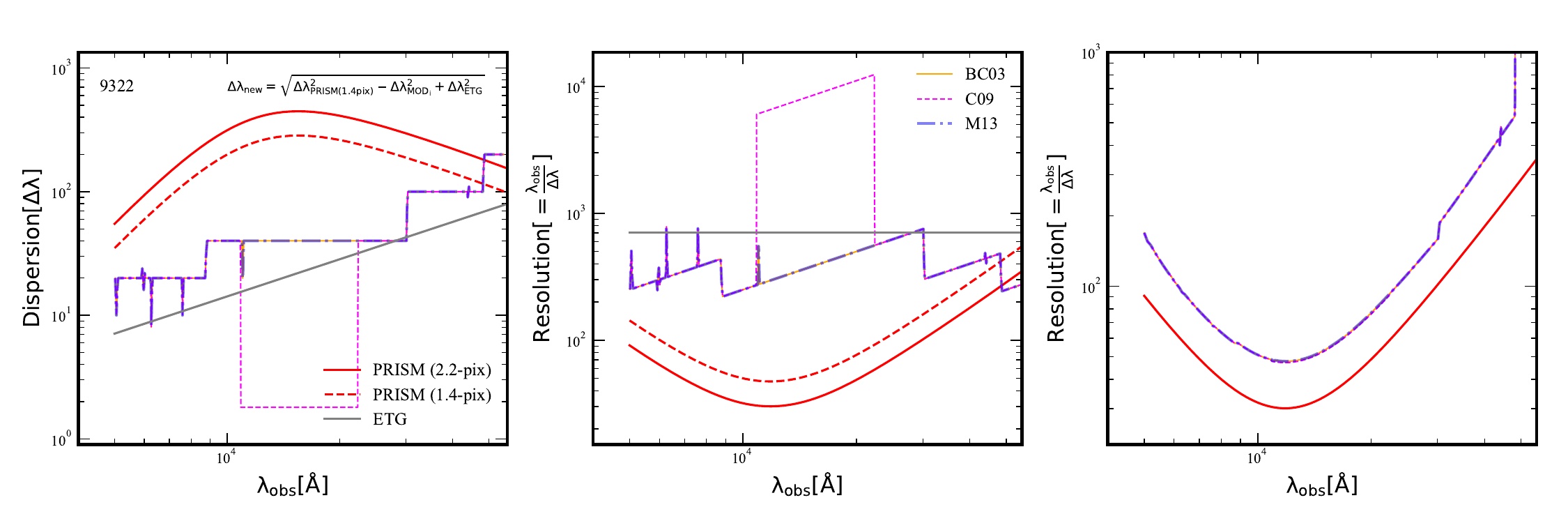}
    \caption{Example of spectral resolution for galaxy ID 9322. From left to right, panels show the dispersion, $\Delta \lambda$ [\AA], the corresponding resolutions for M13 (blue dash-dotted), BC03 (orange solid), and C09 (magenta dashed), and the effective resolution used to match models to the observed spectra. PRISM instrumental resolutions are shown by red lines for 2.2-pixel (pre-launch JWST resolution) and 1.4-pixel (adopted here) shutter elements. The intrinsic velocity dispersion, inferred from local ETGs of similar stellar mass \citep{Thomas+05}, is shown in gray. }
    \label{ext_figC:res_ex}
\end{figure*}

\section{Spectral resolution} \label{AppendixC}
Since different SPS models have different spectral sampling, the final resolution used in the spectral fitting must be adjusted for each model. Following \cite{Lu+25}, we adopt the same methodological approach to account for the combined effects of the instrument, the intrinsic galaxy velocity dispersion, and the wavelength sampling of a specific model on the final resolution. An example for galaxy ID=9322 is shown in Figure~\ref{ext_figC:res_ex}, where a conservative factor of $\times$1.4 (dashed line) is adopted to rescale the in-flight PRISM resolution.

The intrinsic dispersion of each early-type galaxy (ETG) is estimated from the local stellar mass–velocity dispersion (i.e., M-$\sigma$) relation \citep{Thomas+05}. Following \cite{Lu+25}, the velocity dispersion $\sigma$ is converted to the full width at half maximum (i.e., FWHM =$\Delta \lambda$) via 1/R = $\Delta \lambda$/$\lambda$ =v(rest, FWHM)/c. The effect of the specific SPS model sampling is incorporated using
\begin{equation}
    \Delta \lambda_{\rm new} = \sqrt{\Delta^2_{\rm PIRSM (1.4 pix)} - \Delta^2_{\rm MOD_i} + \Delta^2_{\rm ETG}},
\end{equation}
where $i$ refers to the individual SPS models used in this work: BC03 (orange), C09 (magenta), and M13 (blue), as shown in Figure~\ref{ext_figC:res_ex}. Note that at observed-frame wavelengths beyond $\sim$4$\mu$m, the effective resolution is limited by the resolution of the SPS models.

\section{Spectral fit and model setup } \label{AppendixD}
To derive the physical properties of each QG, we adopted a custom IDL code \citep{Gobat+12}, as in \cite{Lu+25}, to fit the aperture-corrected 1D spectra, accounting for the final resolution of each SPS model. The spectral fitting configuration follows \cite{Lu+25}, including a delayed exponentially declining SFH ($\propto (t/\tau^2)e^{-t/\tau}$, i.e., delayed-$\tau$ model), the \cite{Calzetti+2000} attenuation law, and the chosen IMF. \cite{Lu+25} have confirmed that adopting different dust laws does not significantly affect the derived best-fit physical parameters, { consistent with our robustness tests presented in Appendix~\ref{AppendixE.2}.}
For each galaxy, the fit is first explored over a wide, coarse grid and repeated several times to identify a narrow convergence range. The final fit adopts the SSP grids shown in Table~\ref{ex_tabG:SSP_grid}, { similar to those in} Supplementary Table 2 of \cite{Lu+25}, including age, metallicity, $\tau$, and redshift ($\Delta z = 0.001$). Figure~\ref{ext_figD:SED_ex} shows the best spectral fits for galaxy ID=9322 using the M13 (left), BC03 (middle), and C09 (right) models.

The best-fit physical properties derived for each galaxy using a specific SPS model include the stellar mass ($\log M_*/M_\odot$, assuming a \citealt{Chabrier+03} IMF), mass-weighted age ($t$), metallicity ($Z/Z_\odot$), and dust attenuation (A$_{\rm v}$). Table~\ref{ext_tabD:list} lists the best-fit results obtained with the M13 model. The 3$\sigma$ uncertainties are estimated from the $\chi^2$ distribution. 
{ All best-fit ages remain below the age of the Universe at their corresponding redshifts.}

\setcounter{figure}{0} 
\renewcommand{\thefigure}{D\arabic{figure}}
\begin{figure*}[htbp]
    \centering
    \includegraphics[width=0.9\linewidth]{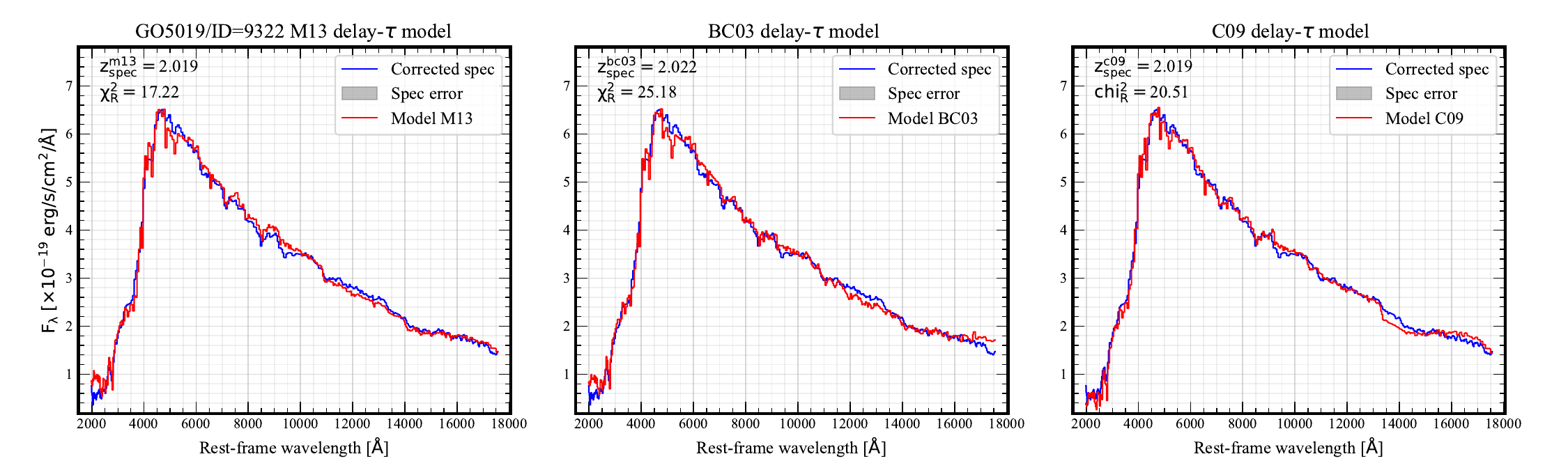}
    \caption{
    Example spectral fits for galaxy ID = 9322 using the M13 (left), BC03 (middle), and C09 (right) models. The aperture-corrected observed spectrum is shown as a blue histogram, with uncertainties indicated by the gray shaded region, while the best-fit model is overplotted in red. The spectroscopic redshift inferred from the best-fit delayed-$\tau$ model, together with the reduced Chi-square ($\chi^2_R$), are listed in the upper-left corner of each panel.}
    \label{ext_figD:SED_ex}
\end{figure*}

\section{Robustness tests} \label{AppendixE}
\subsection{Additional SPS models} \label{AppendixE.1}
{ As an additional robustness test, we repeated the spectral fitting using the CB07 models (https://www.bruzual.org/cb07/) with the same fitting grid and assumptions adopted in \cite{Lu+25}. The inferred stellar population parameters are statistically consistent with those obtained from BC03, while the CB07 models generally yield higher $\chi^2_{\rm R}$ values. The relative ranking of the SPS models remains unchanged, confirming that our main conclusions are robust to the inclusion of this additional SPS model.}

\subsection{Dust Attenuation Assumptions} \label{AppendixE.2}
{ We repeated the spectral fitting under two alternative dust assumptions: adopting the Milky Way extinction law \citep{Fitzpatrick+1999} and assuming no dust attenuation ($A_{\rm V}=0$) for all galaxies. In both cases, the inferred stellar population parameters and the relative ranking of the SPS models remain unchanged. The only noticeable difference is a modest increase in the $\chi^2_{\rm R}$ values for BC03 and C09, whereas the M13 fits are almost unaffected. These tests confirm that our main conclusions are robust against the adopted dust attenuation prescription.}

\subsection{Independent spectral fitting with STARLIGHT} \label{AppendixE.3}
{ We repeated the spectral fitting using the independent spectral fitting code STARLIGHT \citep{Fernandes+05}. The derived ages, stellar masses, and relative performance of the SPS models (i.e., M13, BC03, C09, and CB07) are fully consistent with those obtained with our default fitting code. We therefore conclude that our main results are insensitive to the choice of fitting software.}

\setcounter{equation}{0}
\renewcommand{\theequation}{F\arabic{equation}}
\section{Definitions of Spectral Feature Strengths} \label{AppendixF}
The H-band bump index, I(H-bump), sensitive to the presence of cool evolved stars, reflects the combination of the H$^{-}$ opacity peak at 1.6 $\mu$m and adjacent H$_2$O absorption bands around 1.4 and 1.9 $\mu$m. Following \cite{Lu+25}, we adopt the definition of \cite{Verro+22}:
\begin{equation} 
\rm I(H{\text -}bump) = -2.5 \log [ (\frac{1}{\lambda_2 - \lambda_1}) \int_{\lambda_1}^{\lambda_2} \frac{F_\lambda}{F_{cont}} d\lambda ], 
\label{eq: H-bump} 
\end{equation}
where the pseudo-continuum flux, $F_{\rm cont}$, is estimated from a linear fit to broader continuum windows at rest-frame 1.4-1.5 and 1.8-2.0 $\mu$m, rather than connecting the midpoints of narrower windows (1.45–1.47, and 1.765–1.78 $\mu$m) as in previous work. The integration limits are correspondingly extended to $\lambda_1 = 1.5$ $\mu$m and $\lambda_2 = 1.8$ $\mu$m (cf. 1.61-1.67 $\mu$m), reducing sensitivity to local fluctuations and improving robustness in the stacked spectra.

The CN strength indices at 1.1, 0.92, and 0.78 $\mu$m are quantified by extending the CN1.1 definition of \cite{Lu+25} to other prominent CN edge features:
\begin{equation}
\rm I(CN_{\rm edge}) = -2.5 \log \langle F'{\rm cont} / F\lambda \rangle,
\label{eq:CN_edge}
\end{equation}
where the pseudo-continuum, $F'_{\rm cont}$, is determined from a linear fit to a continuum window blueward of the CN edge. For CN1.1, the blueward window spans 1.06-1.092 $\mu$m ($\sim$160-260\AA), allowing for shifts in the edge during stacking due to wavelength-dependent effects, redshift uncertainties, or instrumental resolution. For CN0.92 and CN0.78, the corresponding blueward windows are $\sim$0.86-0.92 $\mu$m ($\sim$200-350\AA) and $\sim$0.73-0.76 $\mu$m ($\sim$200-300\AA), respectively. A linear slope is allowed in the continuum fit to distinguish genuine edge features from a smoothly varying continuum but if the fitted slope is positive, it is set to zero to avoid artificially enhancing the index. The pseudo-continuum is then extrapolated to the red side of the CN edge, where the flux is integrated to compute the index. The redward windows are $\sim$1.10-1.12 $\mu$m for CN1.1, $\sim$0.93-0.96 $\mu$m for CN0.92, and $\sim$0.79-0.81 $\mu$m for CN0.78.

\section{The spectral fits of stacked spectrum} \label{AppendixG}
The spectral fits to the stacked spectra were performed using the same delayed-$\tau$ SFH and model configuration as for the individual galaxies (see Table~\ref{ex_tabG:SSP_grid}). The input spectral resolution for the stacks was determined during the stacking process (Section~\ref{sec4.2.1}). The best-fit results for stacks remain robust when adopting alternative SFHs (e.g., exponentially declining SFH or non-parametric SFH) and different dust attenuation laws (e.g., Milky Way-type extinction curves; \citealt{Fitzpatrick+1999}).
{ We further tested the potential impact of the stacking scheme by defining the age bins using BC03- and C09-derived ages instead of M13 ages. The resulting stacked spectra yield the same conclusions, with M13 remaining the preferred SPS model.}

Because only three { truly young} galaxies fall in the $t \le 0.4$ Gyr { where TP-AGB contributions are not dominant}, masking emission lines and shifting to the rest frame prior to stacking results in limited NIR coverage with numerous absorption features (see Figure~\ref{ext_figG1:stack_fit_lt0p4}). 
A larger sample is required to confirm these results and to identify the NIR features robustly in future work. Here, we present only the best-fitting models to the stacked spectrum in this younger bin, after masking the $\sim$2000 \AA\ unidentified features, which we will investigate in future work.

{ Within the TP-AGB-dominated subsample, galaxies are further divided into two bins of comparable size according to stellar mass (left panels), mass-weighted age (middle panels), and dust attenuation (right panels). Figure~\ref{ext_figG2:stack_2bins} shows the corresponding stacked spectra (colored curves) together with the individual spectra (gray curves), compared to the spectrum (purple curve) of D36123 from \cite{Lu+25}.}

\setcounter{figure}{0} 
\renewcommand{\thefigure}{G\arabic{figure}}
\begin{figure*}[htbp]
    \centering
    \includegraphics[width=1\linewidth]{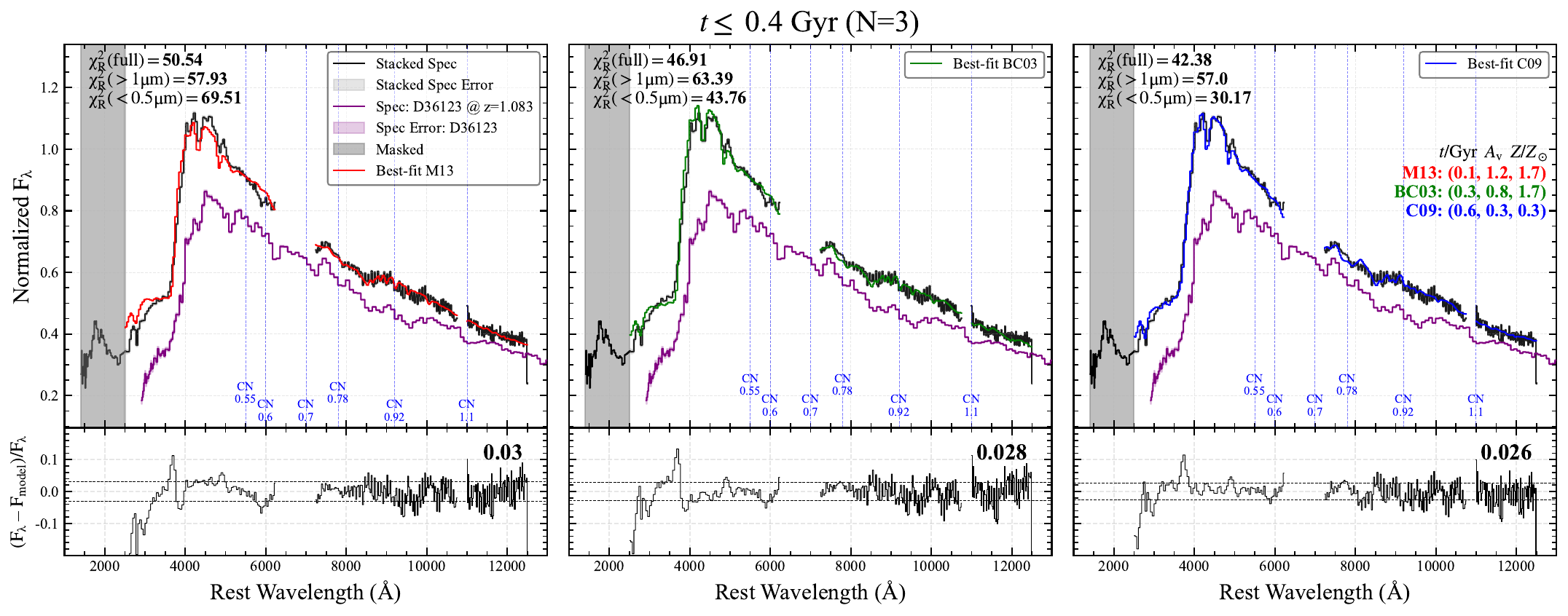}
    \caption{Best-fit models for the stacked spectrum with $t \le 0.4$ Gyr: M13 (left, red), BC03 (middle, green), and C09 (right, blue). The masked region is shown in a grey shaded region. Other symbols and annotations are as in Figure~\ref{fig4:stack_fit}.}
    \label{ext_figG1:stack_fit_lt0p4}
\end{figure*}

\begin{figure*}[htpb]
    \centering
    \includegraphics[width=0.9\linewidth]{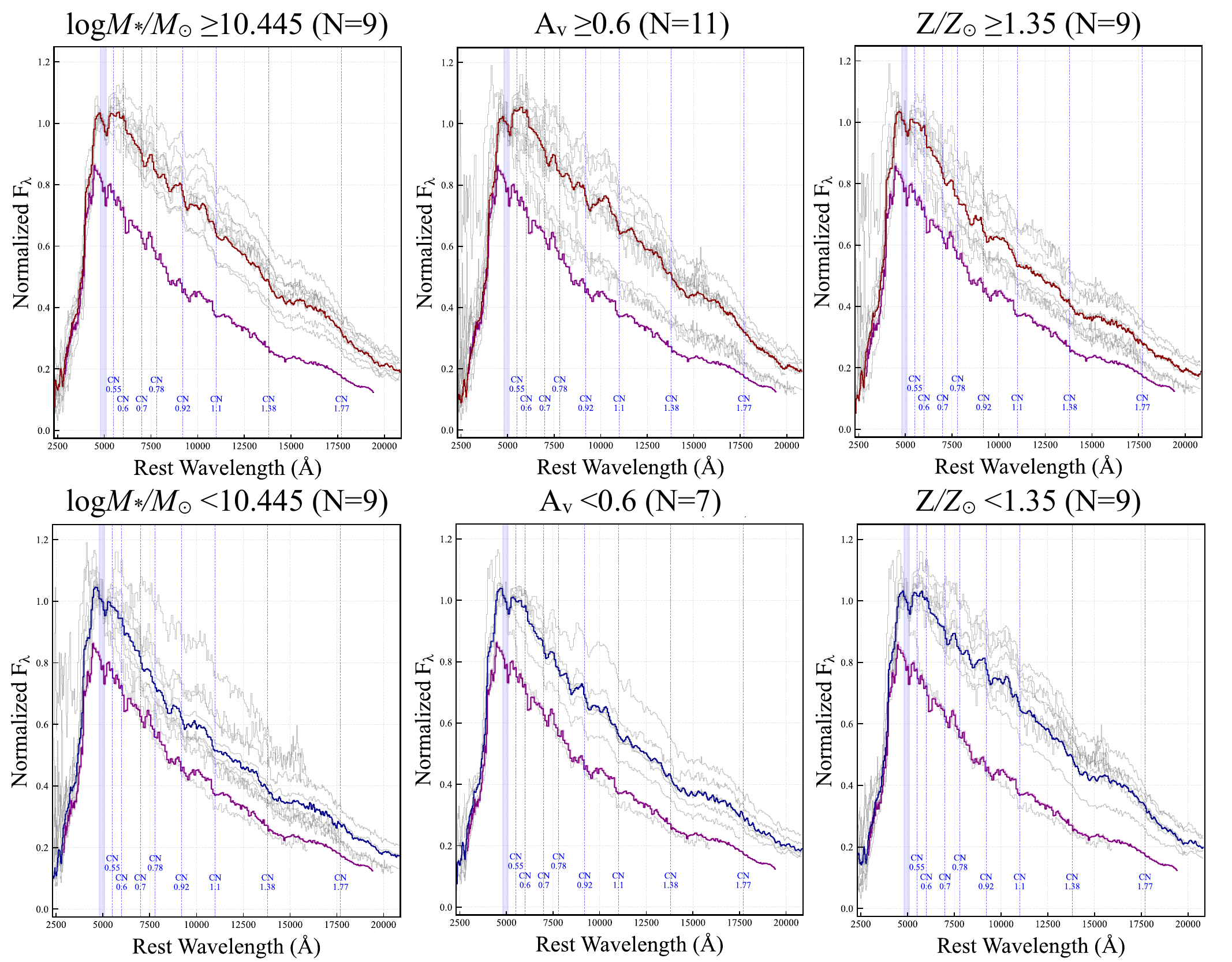}
    \caption{Stacked spectra for galaxies with mass-weighted ages $t=0.4$-$1.8$ Gyr, as in Figure~\ref{fig4:stack_fit}, divided by three selection criteria: stellar mass ($\log M_*/M_\odot$, left), dust attenuation ($A_{\rm v}$, middle), and metallicity ($Z/Z_\odot$, right). High (top) and low (bottom) subsamples are shown as red and blue curves, respectively. { The spectrum of QG D36123 \citep{Lu+25} is shown in purple for comparison}. Thin gray curves denote the individual spectra included in the stacks. Selection criteria and sample sizes are labeled in each panel.}
    \label{ext_figG2:stack_2bins}
\end{figure*}

\setcounter{table}{0} 
\renewcommand{\thetable}{D\arabic{table}}
\begin{deluxetable*}{lllc}[htbp]
\digitalasset
\tablewidth{0pt}
\tablecaption{Parametric grids for the spectral fit \label{ex_tabG:SSP_grid}}
\tablehead{\colhead{Parameters [units]} & \colhead{Min}  & \colhead{Max} & \colhead{Step}}
\startdata
$t^{\dagger}$ [Gyr] &   0.1 & 3.0 & 0.1 \\
& 3.0 & 3.5 & 0.5 \\ 
 \hline
    &   \multicolumn{3}{l}{0.001, 0.004, 0.009, 0.05} \\
$\tau$ [Gyr] & 0.1 & 1.0 & 0.1 \\
    &    1.0 & 3.0 & 1.0 \\
 \hline
$\rm A_{v}$ [mag] & 0 & 2.1 & 0.1 (Calzetti) \\
 \hline  
  & 0.3 & 2.2 & 0.1 (M13)\\
$Z/Z_\odot$  &  0.3 & 2.5 & 0.1 (BC03)\\
  &  0.3 & 2.0 & 0.1 (C09) \\
\enddata
\tablecomments{
Superscript $^{\dagger}$ indicates the time since the onset of star formation. The solar metallicity $Z_\odot=$0.02 is adopted. }
\end{deluxetable*}

\begin{deluxetable*}{llllllllllll}
\digitalasset
\tablewidth{0pt}
\tablecaption{The best-fit physical parameters of 27 QG sample derived by the M13 model \label{ext_tabD:list}}
\tablehead{
\colhead{No.} & \colhead{ID} & \colhead{R.A.($^{\circ}$, J2000)} & \colhead{Decl.($^{\circ}$, J2000)} & \colhead{SNR$_{\rm mean}$} &\colhead{$z_{\rm spec}$} & \colhead{$\log M_*/M_\odot$} & \colhead{Age $t$ [Gyr]} & \colhead{$Z/Z\odot$} &\colhead{A$_{\rm v}$} &\colhead{Program} &\colhead{$z^{\rm adopted}_{\rm spec}$}
}
\startdata
1 & 8502 & 214.974477 & 52.879579 & 92.4 & 2.415$^{+0.001}_{-0.002}$ & 10.575$^{+0.001}_{-0.002}$ & 0.399$^{+0.006}_{-0.005}$ & 1.000$^{+0.038}_{-0.001}$ & 0.800$^{+0.065}_{-0.001}$  & GO-5019 & 2.415\\
2 & 9322 & 214.953564 & 52.868222 & 105.0 & 2.019$^{+0.001}_{-0.001}$ & 10.441$^{+0.001}_{-0.001}$ & 0.700$^{+0.016}_{-0.011}$ & 1.700$^{+0.008}_{-0.078}$ & 0.200$^{+0.057}_{-0.022}$ & GO-5019 & 2.019\\
3 & 10894 & 214.970148 & 52.887768 & 139.5 & 1.357$^{+0.001}_{-0.001}$ & 10.547$^{+0.001}_{-0.001}$ & 1.303$^{+0.028}_{-0.006}$ & 1.200$^{+0.001}_{-0.094}$ & 0.300$^{+0.006}_{-0.028}$ & GO-5019 & 1.357\\
4 & 12809 & 214.912481 & 52.857014 & 161.8 & 1.541$^{+0.001}_{-0.001}$ & 10.557$^{+0.001}_{-0.001}$ & 0.692$^{+0.043}_{-0.001}$ & 1.500$^{+0.040}_{-0.003}$ & 0.300$^{+0.096}_{-0.002}$ & GO-5019 & 1.541\\
5 & 13095 & 214.954077 & 52.888714 & 85.5 & 1.011$^{+0.001}_{-0.001}$ & 10.019$^{+0.001}_{-0.007}$ & 2.523$^{+0.001}_{-0.148}$ & 1.800$^{+0.054}_{-0.003}$ & 0.500$^{+0.027}_{-0.014}$ & GO-5019 & 1.016\\
6 & 13896 & 214.909352 & 52.860793 & 103.6 & 1.084$^{+0.001}_{-0.001}$ & 10.199$^{+0.002}_{-0.001}$ & 1.859$^{+0.028}_{-0.017}$ & 0.900$^{+0.002}_{-0.079}$ & 0.700$^{+0.036}_{-0.005}$ & GO-5019 & 1.180\\
7 & 15675 & 214.899696 & 52.863394 & 80.3 & 1.471$^{+0.001}_{-0.001}$ & 10.444$^{+0.004}_{-0.001}$ & 0.505$^{+0.008}_{-0.013}$ & 1.600$^{+0.007}_{-0.081}$ & 0.700$^{+0.012}_{-0.004}$ & GO-5019 & 1.483\\
8 & 17635 & 214.931706 & 52.895090 & 136.3 & 1.201$^{+0.001}_{-0.001}$ & 10.410$^{+0.001}_{-0.001}$ & 0.701$^{+0.092}_{-0.001}$ & 1.900$^{+0.072}_{-0.077}$ & 0.500$^{+0.001}_{-0.079}$ & GO-5019 & 1.202\\
9 & 20150 & 214.880941 & 52.873604 & 118.0 & 1.539$^{+0.001}_{-0.001}$ & 10.446$^{+0.001}_{-0.001}$ & 0.600$^{+0.003}_{-0.060}$ & 1.100$^{+0.085}_{-0.001}$ & 0.900$^{+0.030}_{-0.001}$ & GO-5019 & 1.532\\
10 & 21384 & 214.874647 & 52.874790 & 139.4 & 1.542$^{+0.001}_{-0.001}$ & 10.450$^{+0.001}_{-0.003}$ & 0.602$^{+0.002}_{-0.060}$ & 1.300$^{+0.097}_{-0.002}$ & 0.500$^{+0.004}_{-0.002}$ & GO-5019 & 1.525\\
11 & 21754 & 214.876885 & 52.877214 & 102.7 & 1.531$^{+0.001}_{-0.001}$ & 10.661$^{+0.003}_{-0.001}$ & 0.602$^{+0.001}_{-0.038}$ & 1.100$^{+0.062}_{-0.001}$ & 1.100$^{+0.046}_{-0.001}$ & GO-5019 & 1.536\\
12 & 21909 & 214.888879 & 52.883553 & 128.7 & 1.037$^{+0.001}_{-0.001}$ & 10.861$^{+0.003}_{-0.001}$ & 2.045$^{+0.001}_{-0.066}$ & 2.000$^{+0.006}_{-0.024}$ & 0.600$^{+0.056}_{-0.008}$  & GO-5019 & 1.065 \\
13 & 9234 & 214.971565 & 52.881214 & 47.4 & 1.087$^{+0.001}_{-0.003}$ & 9.718$^{+0.002}_{-0.001}$ & 2.705$^{+0.004}_{-0.057}$ & 0.900$^{+0.002}_{-0.051}$ & 0.600$^{+0.002}_{-0.091}$ & GO-5019 & 1.380 \\
14 & 14727 & 214.895616 & 52.856494 & 95.0 & 3.240$^{+0.001}_{-0.001}$ & 10.750$^{+0.001}_{-0.001}$ & 0.312$^{+0.003}_{-0.028}$ & 0.400$^{+0.001}_{-0.078}$ & 0.900$^{+0.018}_{-0.005}$ & GO-5019 & 3.240\\
15 & 22897 & 214.870521 & 52.880737 & 43.4 & 1.060$^{+0.001}_{-0.001}$ & 9.359$^{+0.001}_{-0.005}$ & 1.952$^{+0.035}_{-0.120}$ & 0.800$^{+0.014}_{-0.062}$ & 0.200$^{+0.016}_{-0.051}$ & GO-5019 & 1.065 \\
\hline
16 & 24686 & 215.243822 & 53.039733 & 66.8 & 1.175$^{+0.001}_{-0.001}$ & 10.305$^{+0.004}_{-0.001}$ & 0.392$^{+0.079}_{-0.002}$ & 1.700$^{+0.017}_{-0.102}$ & 0.900$^{+0.009}_{-0.052}$ & CEERS & 1.175\\
17 & 25021 & 214.951722 & 52.838717 & 71.0 & 1.356$^{+0.001}_{-0.001}$ & 10.276$^{+0.004}_{-0.001}$ & 0.782$^{+0.062}_{-0.007}$ & 1.800$^{+0.046}_{-0.001}$ & 0.500$^{+0.018}_{-0.041}$ & CEERS & 1.356\\
18 & 26746 & 215.173188 & 53.019868 & 75.0 & 1.141$^{+0.001}_{-0.001}$ & 10.594$^{+0.001}_{-0.001}$ & 0.582$^{+0.001}_{-0.120}$ & 1.200$^{+0.002}_{-0.003}$ & 1.000$^{+0.047}_{-0.002}$ & CEERS & 1.168\\
19 & 4458 & 215.013779 & 52.959092 & 18.1 & 1.576$^{+0.006}_{-0.004}$ & 10.235$^{+0.009}_{-0.008}$ & 0.505$^{+0.029}_{-0.202}$ & 1.700$^{+0.124}_{-0.126}$ & 0.600$^{+0.051}_{-0.023}$ & CEERS & 1.576\\
20 & 5195 & 215.084873 & 53.009733 & 70.5 & 1.648$^{+0.001}_{-0.001}$ & 10.831$^{+0.001}_{-0.003}$ & 0.701$^{+0.008}_{-0.029}$ & 1.200$^{+0.005}_{-0.094}$ & 0.600$^{+0.016}_{-0.080}$ & CEERS & 1.648\\
21 & 8514 & 215.184599 & 53.031394 & 40.2 & 2.157$^{+0.001}_{-0.001}$ & 10.644$^{+0.001}_{-0.003}$ & 0.913$^{+0.065}_{-0.015}$ & 1.800$^{+0.010}_{-0.077}$ & 0.700$^{+0.024}_{-0.050}$ & CEERS & 2.157\\
22 & 2989 & 215.122707 & 53.009839 & 89.6 & 1.093$^{+0.001}_{-0.001}$ & 10.687$^{+0.003}_{-0.001}$ & 1.382$^{+0.085}_{-0.005}$ & 2.100$^{+0.039}_{-0.003}$ & 0.600$^{+0.005}_{-0.056}$ & CEERS & 1.119\\
23 & 9148 & 214.866480 & 52.852697 & 38.1 & 2.292$^{+0.001}_{-0.001}$ & 10.380$^{+0.004}_{-0.001}$ & 0.692$^{+0.099}_{-0.012}$ & 1.300$^{+0.065}_{-0.034}$ & 0.600$^{+0.089}_{-0.044}$ & CEERS & 2.292 \\
24 & 9411 & 215.069301 & 53.009099 & 17.5 & 2.350$^{+0.005}_{-0.007}$ & 10.142$^{+0.024}_{-0.015}$ & 0.779$^{+0.100}_{-0.063}$ & 0.900$^{+0.020}_{-0.103}$ & 1.100$^{+0.112}_{-0.132}$ & CEERS & 2.346\\
25 & 6283 & 215.134876 & 52.970080 & 17.3 & 1.955$^{+0.011}_{-0.007}$ & 10.096$^{+0.018}_{-0.025}$ & 0.543$^{+0.078}_{-0.136}$ & 1.800$^{+0.400}_{-0.432}$ & 1.200$^{+0.185}_{-0.079}$ & CEERS & 1.955 \\
26 & 5244 & 214.905563 & 52.817754 & 23.9 & 2.301$^{+0.006}_{-0.005}$ & 10.056$^{+0.003}_{-0.003}$ & 0.482$^{+0.013}_{-0.068}$ & 0.500$^{+0.069}_{-0.002}$ & 0.100$^{+0.073}_{-0.001}$ & CEERS & 2.301\\
27 & 2962 & 215.163246 & 53.031516 & 49.9 & 1.269$^{+0.001}_{-0.001}$ & 10.285$^{+0.015}_{-0.010}$ & 2.300$^{+0.191}_{-0.070}$ & 1.300$^{+0.013}_{-0.053}$ & 0.000$^{+0.093}_{-0.000}$ & CEERS & 1.269 \\
\enddata
\tablecomments{SNR$_{\rm mean}$ denotes the average signal-to-noise ratio of each spectrum. All listed physical properties and their 3$\sigma$ uncertainties are derived from the best-fit M13 results. Stellar masses ($\log M_*/M_\odot$) are based on the \cite{Chabrier+03} IMF. $z^{\rm adopted}_{\rm spec}$ denotes the redshifts adopted for stacking. }
\end{deluxetable*}

%% For this sample we use BibTeX plus aasjournalv7.bst to generate the

\bibliography{sample701}{}
\bibliographystyle{aasjournalv7}



\end{document}